# Biased Structural Fluctuations due to Electron Wind Force

O. Bondarchuk*, W.G. Cullen, M. Degawa and Ellen D. Williams
Department of Physics
University of Maryland at College Park
College Park, MD 20742-4111

T. Bole and P.J. Rous
Department of Physics
University of Maryland Baltimore County
1000 Hilltop Circle
Baltimore, MD    21250

Direct correlation between temporal structural fluctuations and electron wind force is demonstrated, for the first time, by STM imaging and analysis of atomically-resolved motion on a thin film surface under large applied current ($10^5$ A/cm$^2$).  The magnitude of the momentum transfer between current carriers and atoms in the fluctuating structure is at least 5x to 15x (± one sigma range) larger than for freely diffusing adatoms. The corresponding changes in surface resistivity will contribute significant fluctuation signature to nanoscale electronic properties.

PACS #s:
| | |
|---|---|
| 73.63.-b | Electronic transport in nanoscale materials and structures |
| 73.25.+i | Surface conductivity and carrier phenomena |
| 73.50.Td | Noise processes and phenomena |
| 68.37.Ef | Scanning tunneling microscopy |
| 68.55.-a | Thin film structure and morphology |

*  Present address: Chemical Physics Department, Fritz-Haber-Insitut der Max- Planck - Gesellschaft, Faradayweg 4-6, Berlin, 14195 Germany

**Introduction.**

Due to the size-scaling of fluctuations, the effects of statistical mechanics will be very different at the nanoscale than for macroscopic systems. The effects of nanoscale thermal fluctuations will impact molecular electronic and nanoelectronic contacts [1-3], device stability [4,5], electromigration [6-14] and noise [15-17]. In this work we quantify the relationship of thermal fluctuations with electrical transport by directly observing step fluctuations at the surface of a current-carrying metal film, as illustrated in Fig. 1. Carrier scattering causes a force due to momentum transfer, known as the electromigration wind force, and corresponding changes in the surface resistivity. By convention, this force is written in terms of an effective valence $Z^*$, and the (macroscopic) applied electric field $E$: $F=Z^*eE$ [18-20]. The momentum transfer force felt by atoms at surfaces depends on the local environment: atoms at step edges, near defects, or freely diffusing at the terrace experience different forces [21-23]. These forces can cause substantial changes in surface morphology [24-27], due to mechanisms similar to those well known in electromigration-induced failure [6, 18-20]. Despite its substantial impact upon the morphological evolution of materials, the electromigration wind force is extremely weak, and detecting its effects have required following changes in structure after long periods of current stressing. Here, we will describe direct observation of the effects of the electromigration force on a time scale of seconds by measuring the nanoscale fluctuations of atomic-layer steps [28] on the surface of a current-carrying metallic thin film.

The fluctuations of a surface step are observed *via* a direct measurement of the position of one element of the step as a function of time, *x(t)*. Near equilibrium, step fluctuations can be well-described using the continuum step model [28-31], which predicts that the time-correlation function grows as a power law for times less than the correlation time. For the system described here, steps on Ag(111), step motion is driven by step edge diffusion (SED) [30, 32, 33], for which the correlation function is:

$$G(t) \equiv \langle [x(t) - x(0)]^2 \rangle = a_x^2 \left(\frac{t}{\tau_4}\right)^{\frac{1}{4}} . \qquad (1)$$

Here $x(t)$ is the position of the step perpendicular to the average step-edge orientation and the average is taken by repeated observations, $a_x$=0.25 nm and $a_y$=0.29 nm are the lattice constants perpendicular and parallel to the step edge. The time characteristic of thermal fluctuations of the step edge, $\tau_4$, is determined by the step stiffness, $\tilde{\beta}$, and the hopping time $\tau_h$ for atomic motion along the step edge:

$$\tau_4 = \left(\frac{\pi}{2\Gamma(3/4)}\right)^4 \left(\frac{\tilde{\beta} a_x}{kT}\right)^3 \left(\frac{a_x}{a_y}\right)^3 \tau_h, \qquad (2)$$

where $\Gamma$ is the Gamma function, and the value of $\tau_h$ has been measured to decrease from ~ 3 μs to ~3 ns between 300K and 460K [33].

Recently the step continuum model has been expanded to include the effect of an electromigration force acting perpendicular to a step that is fluctuating via SED [34]. The correlation function deviates from the equilibrium result as:

$$G(t) = a_x^2 \left(\frac{t}{\tau_4}\right)^{\frac{1}{4}} \int_0^\infty \frac{1}{\left(u^2 \pm \sqrt{t/\tau_{EM}}\right)} \left(1 - e^{-2u^4 \mp 2u^2\sqrt{t/\tau_{EM}}}\right) du \approx a_x^2 \left(\frac{t}{\tau_4}\right)^{\frac{1}{4}} \left[1 \pm 0.3487 \left(\frac{t}{\tau_{EM}}\right)^{1/2}\right] \qquad (3)$$

where the + and – signs correspond to a downhill and uphill direction of force respectively. The presence of the electromigration force gives rise to an additional time constant, $\tau_{EM}$, that represents the time when the time-correlation function begins to deviate significantly from its equilibrium behavior,

$$\tau_{EM} = \frac{(k_B T)(\tilde{\beta} a_x)}{(F a_x)^2}\left(\frac{a_x}{a_y}\right)\tau_h, \qquad (4)$$

Since the electromigration force is weak, $|F a_x| \ll k_B T$ and $|F a_x| \ll \tilde{\beta} a_x$ we anticipate that $\tau_{EM} \gg \tau_h$. The nature of the result indicated by Eq. 3 can be understood by analogy to the Bales-Zangwill `kinetic instability [35]. In both cases, a diffusional bias perpendicular to the step edge results in spontaneous increased deviations from the equilibrium position when the bias favors diffusion in the step-downhill direction. For an up-hill bias, an anomalous straightening of the step edge occurs.

In the remainder of this letter, we exploit this model. Measurement of the correlations of thermal step-edge fluctuations in the presence of current stressing yield the electromigration time constant (Eq. 3) and, as a result, the electromigration force felt by atoms diffusing along the step edge (Eq. 4). This represents a direct measurement of the effective valence of an atom at the surface of a current-carrying solid.

The experimental methods for preparing atomically clean (111)-oriented silver films have been described previously [33, 36]. The films used here were 100 to 200 nm thick and 1-2 mm wide, with micron-scale areas of flat (111)-oriented surface separated by deep pits, which covered about 50% of the film area at the smallest film thickness (100 nm). Atomic cleanliness was confirmed by atomic-resolution STM imaging. Imaging was performed using tunneling conditions of 0.6-0.8 nA and 1V, at a scan rate of ~9 pixels/ms, which are known not to perturb the measured step configurations [30, 32, 33]. After completing the STM measurements on each sample, the sample temperature was measured using a thermocouple brought into direct contact with the film surface [33]. The thermocouple values and the measured hopping time constant [33] were used to determine the sample temperatures.

The temporal evolution of the step fluctuations was observed by repeated STM scans across the step boundaries as shown in Figure 2. The size of each image is 100 nm x 512 scans (56.6 ms/scan) and an electric current of 0.4 A (nominal current density $1\times10^5$ A/cm$^2$) flows through the sample perpendicular to the step edges. The temperature of the current-stressed sample (due to Joule heating) was 370±30 K. By following the motion of the edges of the steps shown in Fig. 2, we simultaneously determine the spatial variations x(t) for up-hill and down-hill steps fluctuating at the same sample temperature under the same current density.

The time correlation functions obtained from the measured x(t) are shown in Fig. 3. The magnitude of the correlation function grows more rapidly for the up-hill current than for the down-hill current, consistent with an electromigration force acting in the same direction as the electron flow in the sample (e.g. $Z^* < 0$, an electron wind force). The data were fit to Eq. 3, with the thermal and electromigration time constants ($\tau_4$ and $\tau_{EM}$ respectively) as the only adjustable parameters. The results of these fits are shown as solid curves in Fig. 3, and demonstrate excellent agreement between the measured correlation functions and those predicted by the Langevin theory (eqn. 4). As can be seen in fig. 4, the two fits yield clear $\chi^2$ minima for $\tau_{EM}$ of 16s and 52s, respectively for the down-hill and up-hill steps. Measurements on two additional steps subject to up-hill current stressing at a higher nominal current density ($J_{nom} = 4\times10^5$ A/cm$^2$) gave fitting results similar to those of Fig. 3, with time constants of 98s (325K) and 32s (350K). The uncertainties in the fit parameters were 15-40%. The correlation function for an unstressed sample measured at 325K is also shown in Fig. 3. There is no minimum in the chi-squared value for the fit as a function of electromigration time constant, with virtually no change in the goodness of fit occurring for values larger than 9000s (see fig. 4). This shows that data is well fit by a single parameter, $\tau_4$, as expected when no electromigration force is acting (i.e. $\tau_{EM} \to \infty$). Similar results were consistently obtained for other steps measured without current stressing.

The electromigration force is found using the measured values of $\tau_{EM}$ in Eqs. 1 and 4, given the step stiffness $\tilde{\beta}$. The stiffness is calculated [37, 38] using an effective kink energy of $\varepsilon = 0.117$ eV [39, 40]. The four measurements of the electromigration time constant yield average values of the force per step-edge atom of $-2.7 \times 10^{-5}$ eV/nm for $J_{nom} = 4 \times 10^5$ A/cm$^2$ (325-350K) and $-9.7 \times 10^{-6}$ eV/nm for $J_{nom} = 1 \times 10^5$ A/cm$^2$ (370K). Thus, within the cumulative uncertainties in the measured forces of $\pm 50\%$, the force increases in direct proportion to the current density.

The measured values of the force can be used to determine the effective valence Z* if the local surface current density and the resistivity in the surface region are known. Since these local quantities cannot be measured directly, we first estimate them using the bulk current densities and the bulk resistivity of Ag, which is approximately $1.8 \times 10^{-6}$ $\Omega$-cm at 325K and $2.2 \times 10^{-6}$ $\Omega$-cm at 370K [41, 42]. The resulting effective valence, obtained using the nominal current density, is Z* = $-(4\pm2) \times 10^2$. The magnitude is substantially larger than the predicted effective valence of an isolated Ag adatom on Ag(111), which is is $Z^* = -19$ [43]. For atoms in a close-packed site along a step edge, with a perpendicular current direction, the direct force per step atom may be as much as 2x higher than the force on an adatom [21, 23], which would yield a predicted valence of Z* ~ -38, still much smaller magnitude than the measured value. A substantial systematic effect can be attributed to the film cross section, because as described earlier, at 100 nm film thickness there are vacancies in the film up to 50% of the surface area. Therefore the bulk current densities may be as much as 2x higher than the nominal values, and as a result the *lower limit* on the effective valence is Z* = $-(2\pm1) \times 10^2$. The remainder of the difference compared with the perpendicular force on a close-packed step-edge atom may arise from the highly kinked environment suggested in Fig. 1. Because diffusion is parallel to the step edge, only the component of the electromigration force tangential to the local step orientation will affect step-edge diffusion. The largest impact of the electromigration force will thus occur for the most highly kinked step regions. There have been no calculations of the electromigration force acting on low-symmetry kink sites. However, such sites have enhanced valence charge density [44], and also present anomalous barriers to step-edge diffusion [45]. Such significant changes in local electronic structure may be reflected in significant changes in the scattering cross section (and thus Z* value). In addition, the geometric effect of the kink configuration is likely to enhance scattering via blocking [21], or constriction-induced enhancement of local current density [46] analogous to current crowding [47, 48].

The forces measured above are related to equal and opposite forces on the charge carriers, which translate into changes in the surface resistivity [22, 43]. This can be evaluated by treating the diffusing step-edge atoms as independent scattering sites of density $n_k = (L_k L_{step})^{-1}$, where $L_k$ is the average distance between diffusing atoms along the step edge and $L_{step}$ is the average distance between steps. Then the change in the surface resistivity $\rho_s$ due to the diffusing step-edge atoms is [43]:

$$L_f \frac{\partial \rho_s}{\partial n_k} = -\frac{1}{en_{Ag}J}\left(F_w^s + \sum_j \delta F_w^j\right), \quad (5)$$

where $L_f$ is the film thickness, $n_{Ag} = 58.5$ nm$^{-3}$ is the bulk carrier density for Ag, $F_w^k$ is the wind force acting per atom (our measured value) and $F_w^j$ are additional changes in force on the carriers due to the perturbation of atomic structure in the immediate vicinity of the step-edge atoms. Using the measured force per atom and the upper limit of the current density yields the component of the surface resistivity due to scattering at the diffusing step-edge atoms alone,

$$L_f \frac{\partial \rho_s^s}{\partial n_k} \geq -\frac{F_w^k}{en_{Ag}J_{max}} \sim (3 \pm 1.5 nm^3)\rho_o,$$ where $\rho_0$ is the bulk value of the Ag resistivity.

The impact of diffusing step-edge atoms on surface resistivity will include the direct-scattering term measured here as well as the perturbed lattice terms ($F_w^k$ in Eq. 5). The latter term may contribute as much as 2/3 of the total resistivity change, thus it is reasonable to expect that scattering at the step-edge will contribute a total change in the surface resistivity

$$L_f \frac{\partial \rho_s}{\partial n_k} \geq (10 \pm 5 nm^3)\rho_o.$$ As an example, for moderate step and kink densities ($L_{step}$ =10 nm and $L_k$ =2 nm) and a very thin film ($L_f$ = 10 nm), the change in surface resistivity due to scattering at the step-kink sites could be as large as 10% of the bulk resistivity. This effect will be significant in nanoelectronic devices carrying large current densities [49, 50].

The present observation of biased temporal fluctuations under current stress is the first direct correlation of nanoscale structural fluctuations with the electromigration wind force. Because of systematic uncertainty in the actual current density, analysis of the results yields a *lower limit* for the magnitude of the electron wind force on the diffusing step-edge atoms. Taking the one-sigma limits on the statistical uncertainties, we find that the value is five to fifteen times larger than that on individual adatoms [43]. Effects on resistivity, noise and electromigration susceptibility in metal nanoelectronic structures will thus be concomitantly higher than would have been expected [51].

**Acknowledgments**

This work was supported by the U.S. Department of Energy Award No. DOE-FG02-01ER45939. We also gratefully acknowledge support and SEF support from the NSF MRSEC under grant DMR 05-20471.

Figure Captions

**Fig. 1:** Schematic illustration of current flow perpendicular to average orientation of steps on the surface. Enhanced scattering from step sites at the surface is suggested by the arrows. The inset illustrates adatoms on the terraces and the kinked (thermally roughened) step edge.

**Fig. 2:** STM data for a current stressed sample. Applied current was 0.4 A and nominal sample cross section 200 nm thick x 2 mm wide. Sample temperature = 380 K. Upper panel – repeated scans across the edges of three steps (pseudo image) (from left to right one downhill step and two uphill steps) show fluctuations of step position in time x(t). Lower panel – height vs. position across the image is consistent with steps a single layer (0.236 nm) high.

**Fig. 3**: Time correlation functions, G(t) in units $\text{Å}^2$, for the step fluctuations (x(t) data) extracted from repeated measurements as shown in Fig. 1 and described in the text. The data for the step-down current (open circles) is the average of 10 separate measurements, and for the step-up current (open squares) is the average is over 9 data sets. Fits to Eq. 4 are shown as the solid lines for each data set. Also shown is the measured correlation function (average of 19 separate measurements) for the unstressed sample (open triangles) fit to a single power law.

**Fig. 4** The reduced chi-squared plotted as a function of $\tau_{EM}$, while holding the value of $\tau_4$ at its optimum value. Solid curve: step-down current. Dot-dashed curve: step-up current. Dashed curve: unstressed data. Note that, for the unstressed data, there is no clear minimum value of $\tau_{EM}$, i.e. a fit to a single parameter ($\tau_4$) suffices to describe the data.

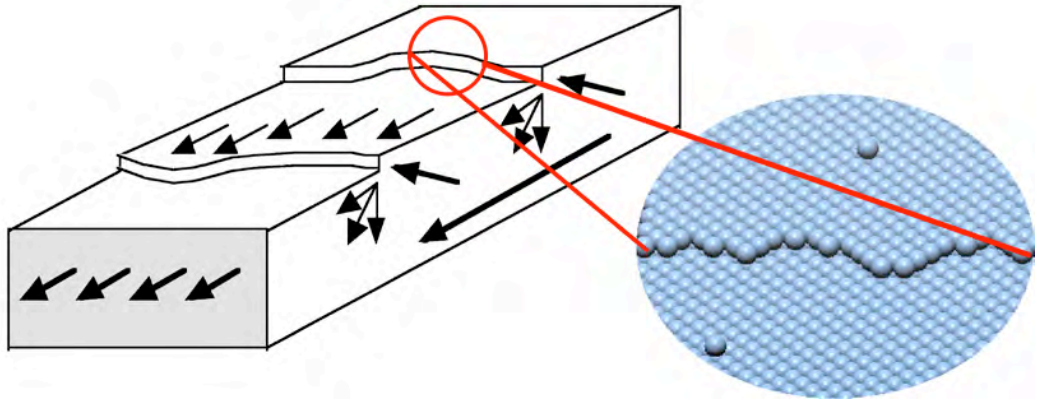

**Fig. 1:** Schematic illustration of current flow perpendicular to average orientation of steps on the surface. Enhanced scattering from step sites at the surface is suggested by the arrows. The inset illustrates adatoms on the terraces and the kinked (thermally roughened) step edge.

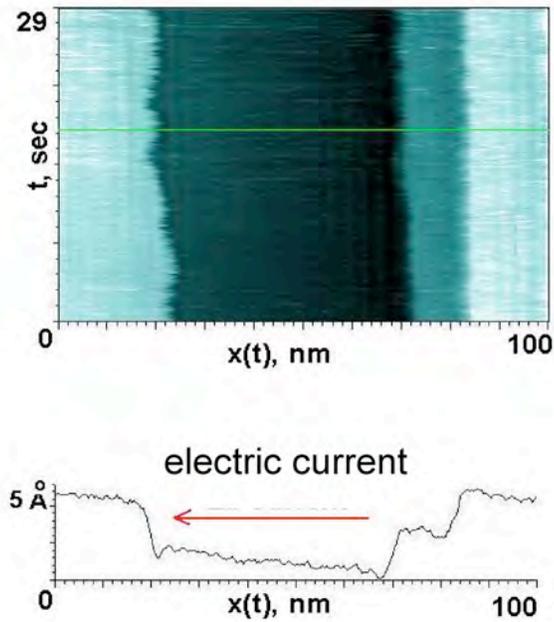

Fig.2: (Color on-line) STM data for a current stressed sample. Applied current was 0.4 A and nominal sample cross section 200 nm thick x 2 mm wide. Upper panel – repeated scans across the edges of three steps (pseudo image) (from left to right one downhill step and two uphill steps) show fluctuations of step position in time x(t). Lower panel – height vs. position across the steps of single layer height (0.236 nm).

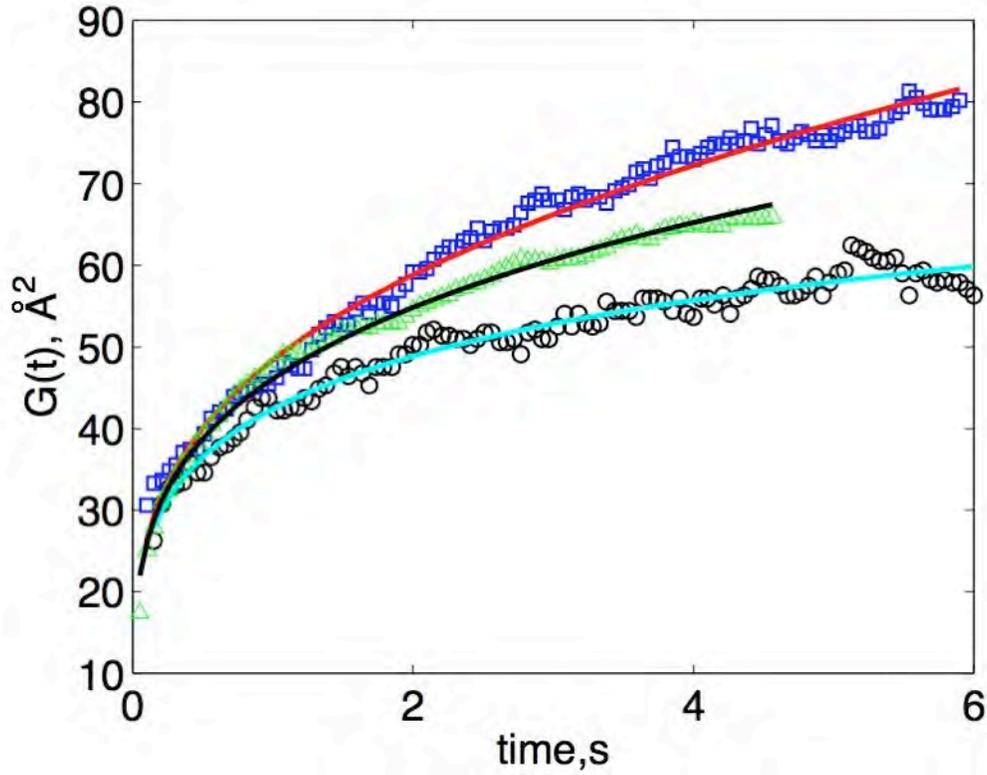

Fig. 3: (Color on-line) Time correlation functions, G(t) for the step fluctuations (x(t) data) extracted from repeated measurements. The data for the step-down current (open circles) is the average over 10 separate measurements, and for the step-up current (open squares) the average is over 9. Fits to Eq. 3 are shown as the solid lines for each data set. Also shown is the measured correlation function (average of 19 separate measurements) for an unstressed sample (open triangles) fit to a single power law.

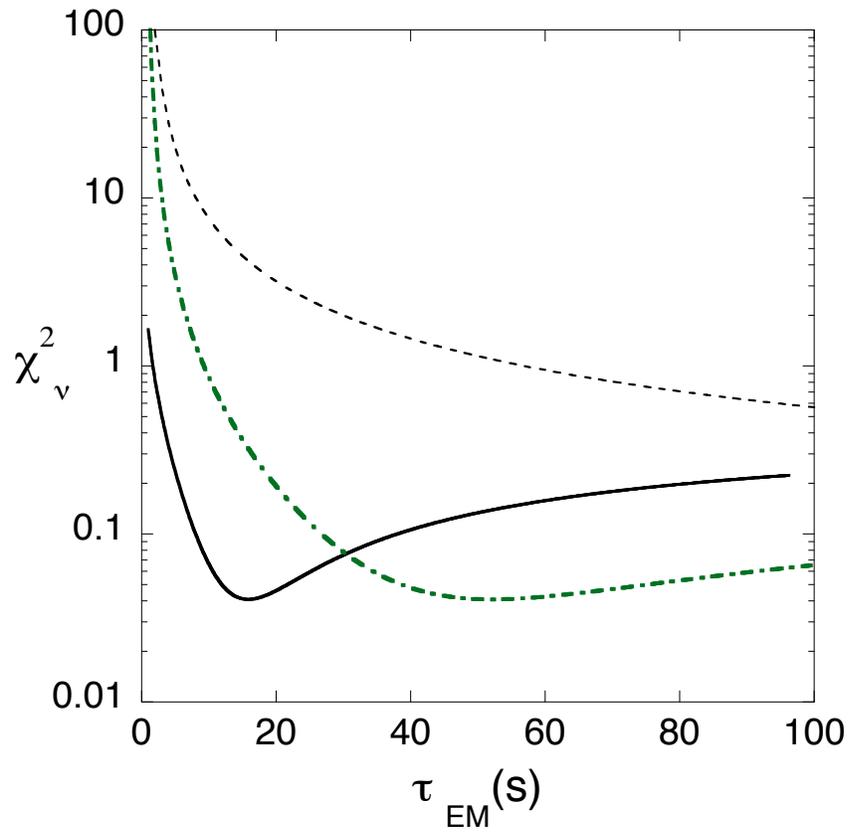

**Fig. 4** (color on-line)The reduced chi-squared plotted as a function of $\tau_{EM}$, while holding the value of $\tau_4$ at its optimum value. Solid curve: step-down current. Dot-dashed curve: step-up current. Dashed curve: unstressed data. Note that, for the unstressed data, there is no clear minimum value of $\tau_{EM}$ that is consistent with the data, i.e. a fit to a single parameter ($\tau_4$) suffices to describe the data.